# RESULTS FROM IMPACT ASSESSMENT ON SOCIETY AND SCIENTISTS OF FRASCATI SCIENZA EUROPEAN RESEARCHERS' NIGHTS IN YEARS 2006 – 2015

S. Arnone[1], F. Bellini[2], M. Faccini[1,3], D. Maselli[1,4], G. Mazzitelli[1,4], I. Paximadas[1], F. Spagnoli[1,4]

[1] *Frascati Scienza (ITALY)*
[2] *Sapienza Università di Roma (ITALY)*
[3] *INAF Osservatorio Astronomico di Roma (ITALY)*
[4] *Laboratori Nazionali di Frascati – Istituto Nazionale di Fisica Nuclere (ITALY)*

## Abstract

The aim of this paper is to present the impact achieved by Frascati Scienza Association on society and research through the European Researchers' Night project funded by the European Commission within the years 2006-2015. The project has been devoted to raise awareness of researchers' work, encourage the dialogue between researchers and citizens and the choice of young people to pursue a career in science.

The first scientific activities and cultural events took shape in 2006, under the coordination of the National Institute for Nuclear Physics (INFN), through the European Researchers' Night project [1], the most important and significant event to promote the role of the researcher and bring people of all ages closer to the scientific world. The positive and successful experience of the first two events, pushed the researchers and citizens of Frascati, where most of Italian research centers and infrastructures are located, to formally associate in the Frascati Scienza in 2008, who started to coordinate the event from 2008. Frascati Scienza was driven by the need to promote educational activities to citizens, young people and schools, in order to involve the general public in science and to bring researchers closer to society.

Every year, the challenge has become ever larger with a progressive increase in exchanging and sharing activities among citizens and researchers, where the latter are dedicated to tell their passion about the scientific world and to transfer their knowledge to "non-scientists". The ten years of the European Researchers' Night experience had an interesting impact on the researchers who have realized that an interaction with the non-scientist people is required to educate not only the society, but also the research world, not used to deal with non-experts.

Since the first edition of the European Researchers' Night, we posed ourselves and to the actors involved several questions about how a science communication event could have an impact on both citizens and researchers to promote the scientific culture and to generate a greater understanding of issues and concepts related to research. How to engage people? How to explain and re-launch the role of researchers? How to investigate the opinions of those who think they should not be involved in research?

The paper describes the impact achieved in each year of European Researchers' Night (ERN) and presents a detailed analysis of the 2014 edition (with 450 ex-ante and 543 ex-post semi-structured questionnaires) and 2015 results (with 304 ex-ante and n. 623 ex-post questionnaires for the quantitative analysis and on the qualitative information collected through the World Café methodology).

## 1 INTRODUCTION

The association has always had the objective to develop a dialogue with the public through a constant exchange of knowledge over time and by encouraging a participatory approach. In this perspective, citizens became valuable partners for researchers on key issues of modern research and on its application. A proper science education helps people to become more responsible and aware of the world around them and to develop a greater understanding about the events of everyday life. The need to revitalize the role of the researcher plays a crucial role in modern society and people have hardly evidence of his work as scientist and how this impacts their life.

The researchers are perceived as distant from the real world, sometimes even as unpredictable and mysterious. The dialogue between researchers and citizens is necessary to understand that science belongs to all and not only to scientists. Hence, the importance of providing an event to give the opportunity to come into direct contact with the researchers, possibly through a fun and participatory methodology is evident. It is very important to involve scientist and non-scientist in multiple ways, so questionnaires over the years have highlighted if people found useful participating to the event, e.g., understanding researchers work or increasing the knowledge of the research role.

Frascati Scienza aims not only to disseminate scientific information, but also to facilitate conversations about science, because a more open science to the public can benefit of more perspectives and discover new collective forms of knowledge.

## 2 METHODOLOGY

The comprehensive survey conducted on the results of the Researchers' Night within the years 2006-2015, was primarily based on a non-probabilistic sample of respondents.

The survey included two types of questionnaires: an ex-ante questionnaire distributed prior to the event, and an ex-post questionnaire administered at the end of event. The number of respondents within all years falls in the range of about 300-900. The distribution has been developed by sending online questionnaires to the subscribers of the newsletter of Frascati Scienza. Data collection took place through the website www.frascatiscienza.it.

Part of the ex-post questionnaires have been collected during the European Researchers' Night, both through the compilation in paper form and directly online through the website and a computer available at the Infopoint.

The main objective of the analysis was to gather useful information to improve future events and/or confirm any expected performance, through the following steps:

- define the target audience reached during the European Researchers' Night
- measure how much the event has affected the awareness on commitment, knowledge and promotion of the researcher's figure
- draft the most possible concrete idea of the image of the researcher
- being able to understand how much difference, between Italy and Europe, is perceived by the public interviewed in relation to employability, consideration given to the research and funding
- analyse how much the role of research in Italy and Europe is considered important
- evaluate how much and which media influenced the dissemination of news and event information
- understanding what was the level of appreciation of the events.

## 3 IMPACT OF ERN ACTIVITIES DEVELOPED BY FRASCATI SCIENZA: 2006 – 2015

With reference to the gender representation, only during the editions 2006 – 2007 of the European Researchers' Night there has been a male representation stronger than women (with a predominance of men of 20 percentage points).

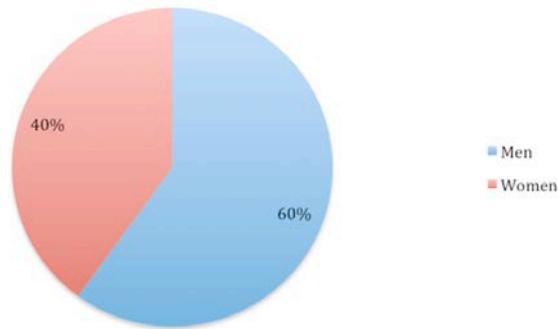

Figure 1 Gender representation in years 2006-2007.

In the following years, until the edition of 2015, the number of women participation tended to be equal to that of men. In 2015 the number of men and women is almost comparable, with approximately 9 percentage points of difference in favor of women representation.

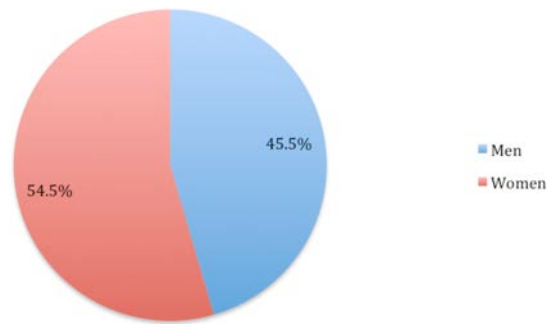

Figure 2 Gender representation in year 2015.

Besides the growing interest of women to research and related activities, it is relevant to note how much the internal composition of age groups has been changed from the editions, reflecting a progressively trend to an increase of participation of young people and adults in 2015, especially considering in the women class (from 10-19 to 40-49).

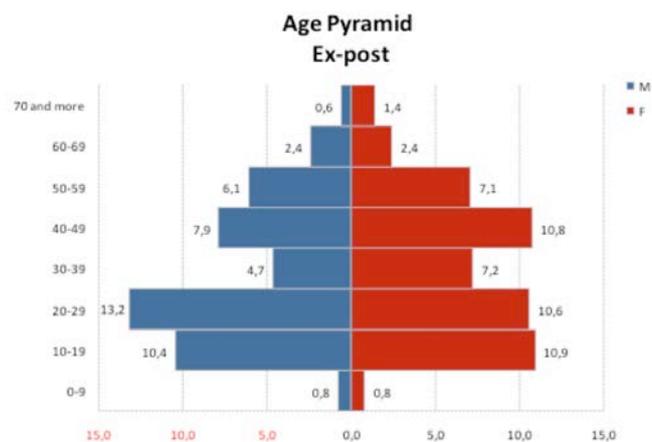

Figure 3 Age Pyramid in year 2015.

The fact that the participants are focused on the younger and central classes of age groups denotes also the importance of specific choices in communication, required by the technological evolution of the media used by the target (mainly internet and social networks). Anyway, the use of both new media and old media (such as outdoor advertising) has developed interesting results since they constituted the key of advertising campaigns and for the events included in the program (for example competitions for schools and young people from 2007-2014, "The researchers show their face" for 2010 edition, the Flashmob of 2011, "Speed date with researchers" in 2012 and 2013).

The young/adult participants of the European Researchers' Night have predominantly a high level of education, as the respondents appear to be mainly *university student* or already have a *degree title* (in the various editions it is possible to state that the percentage value, on average, is approximately of 35% for both variables). Throughout the different editions, in order to investigate the situation of researchers in Italy, the question posed was: *"If you are a researcher, do you feel professionally inserted in Italy?"*. There has been a changing trend in considering the role of the researcher in society. Although the answer *little* is always present, until the Night 2011, it is detectable even a good percentage of those who believed that the researcher was *enough* inserted in Italy (on average about 32%), while from 2012 to 2015, this percentage tends to decrease and the *not at all* answer was predominant. It can be deduced that the perception of the role of the researcher in Italy is even more pessimistic. More than the half of researchers in 2014 (about 59%) answered to feel *little* or *not at all* professionally inserted in Italy.

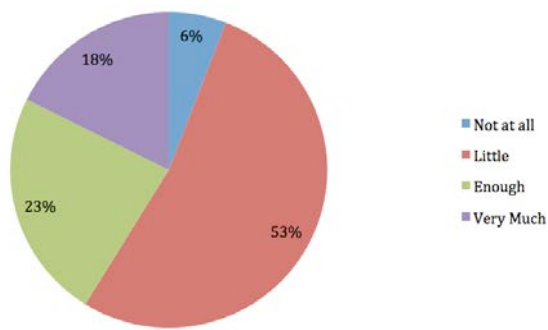

Figure 4 Perception of the role of the researcher in Italy in 2014.

From the edition of 2006 to 2008, most of the people interviewed stated to have become aware of the event especially through *internet, friends/word of mouth and direct contacts with the Research Institutes involved* (on average, respectively, approximately 26% in 2006, 30 % in 2007 and 26% in 2008). From 2009 [2] to 2012 [3] [4], the trend for the variables listed above has not changed. In these years we focused the advertising campaign on the visual identity and we have decided to invest more on *outdoor advertising*, this strategy has shown a positive impact in engaging more the society in the events (on average about 15%). During the same period, the use of internet to reach people (through its various channels: Frascati Scienza website, daily press and radio websites, social networks, etc.) has been on average about 35%. From 2013 to 2015 it is possible to note different data from the last trend and this is probably due to a change in the media plan compared to previous years. This identifies a return to a generally lower use of media, showing a predominance of those who have been reached through *friends/word of mouth* (about more than 50%) and 26% from other sources (in which *internet* is included, about 11%, considerably reduced compared to all other editions).

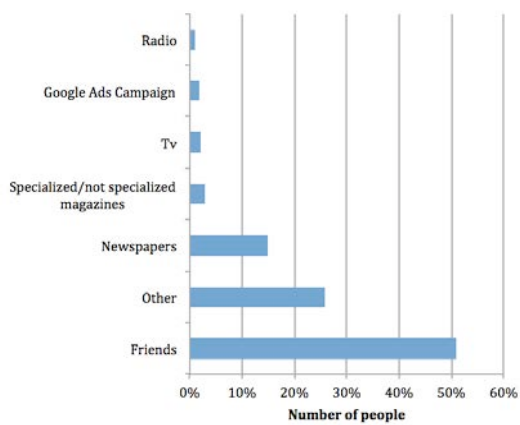

Figure 5 Means of communication from 2013 to 2015.

Throughout the different editions of the European Researchers' Night, on average, 84% of respondents stated that the role of research is *very important for the development of our country and for Europe*. It might be assumed that there is a common direction of Italy and Europe related to the significance of research for the next future.

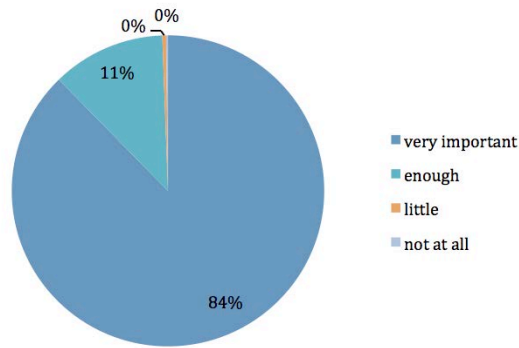

Figure 6 The importance of the role of research for the development of Italy and Europe.

There are enormous differences when we deal with financing of research projects in Italy and Europe. An overwhelming majority of respondents have stated to be aware of *not appropriate financing* of the research in Italy (on average 87%). While, in Europe, the research initiatives are considered *well financed* for about 72% of respondents, on average. It is possible to deduce almost an opposite trend.

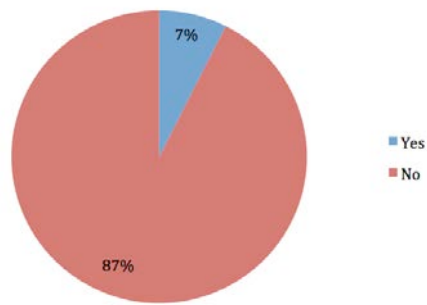

Figure 7 Do you think that in Italy research is adequately funded?

From the event of 2010 to that of 2015, it has been requested a list on which to place the various scientific disciplines that people believed important for the research. For all the years, on average percentage, *Medicine* has been considered as the most relevant (about 41%), followed by *Physics* (about 17%).

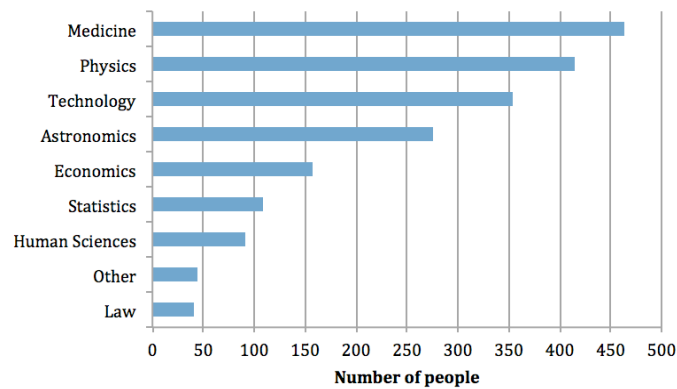

Figure 8 The subject in which you feel the important role of research.

Until the 2009 edition, people had to answer *whether the event had helped to increase knowledge on research*, with positive results (on average, 78%). While, in the editions from 2010 onwards, it was considered necessary to split the question into multiple answers mode, in order to identify more specifically how the event has helped to understand it. Therefore, in the following years, the result is that the event has contributed *to give a clearer image of what research is*, on average, for about 40% of participants until 2015, in which *raise awareness of researchers' work* is considered the most important category (66,3%).

Interestingly, within the editions from 2010 to 2015, the event helped in providing a *clearer understanding of what research is*, indeed the percentage of people supported this increased, starting from 21% in 2010, reaching 56% in 2015. This survey could be interpreted as an increase of researchers' communication effectiveness, adapted to the needs of the target, through more dedicated and interactive events.

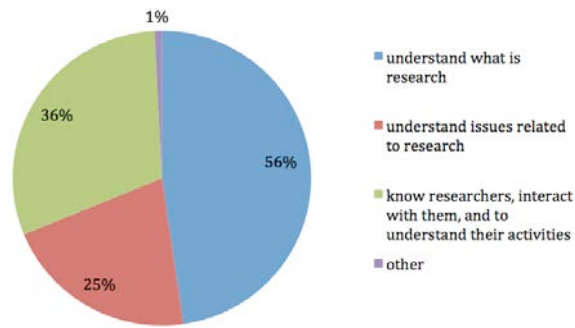

Figure 9 Contribution of the event.

Overall, the people interviewed between 2009 and 2015 believe that *the event has changed their idea of the figure of the researcher in a positive way.* Overall, the respondents in recent years of the event have described the figure of the researcher with the following characteristics: *enough young, certainly useful, dynamic and important, but not rich and not fully integrated into the society as a relevant figure*.

On average, 88% of participants has considered the event *very useful to promote the role of the researcher*. Finally, the *event has been judged positively*, on average, by about 85% of the participants in all the editions and about 98% of respondents has stated that *would like to participate again in an event of this type.*

## 4 ERN 2015 IMPACT AND WORLD CAFÉ RESULTS

To achieve the main goal of evaluating the impact of the "European Researchers' Night of 2015" in the public perception of the researchers and their work, quantitative and qualitative analysis have been realized. A total of 927 questionnaires (304 ex-ante and 623 ex-post) have been collected.

With reference to the questionnaires, the large number of men and women participating in the event is almost comparable, with approximately 9 percentage points of difference in favor of women representation. The highest percentage of those who filled in the ex-ante questionnaire refers to the class of 40-49 years old, followed by 50-59 and 20-29. Unlike the event in 2014, it seems there is no significant representation of the 30-39 age group. The interviewed ex ante population is mostly employee (39%) and University student (15%). Respondents to the ex-post questionnaires are mainly representing the 20-29 years old group (24%), 10-19 years old (22%) and 40-49 years old (19%). There is still a lower percentage of those who fall in the range 30-39 years old, but the percentage of younger classes increases. This has also happened during the ex-post analysis of the 2014 event. Those who have attended the events have mostly an University degree (36%) or a High school Diploma (26%). According to the public interviewed after the event, Friends/Word of mouth has been the medium most used for promoting the ERN (28%), then Internet (18%) and direct contacts with the Research Entities Involved (16%). From Night 2014, Friends/Word of mouth is the prevailing medium. Half of the respondents claim to have attended the previous editions of the ERN. Both in Italy and in Europe, most of the people interviewed consider the role of research for countries' development very important. Even in the Night 2015, as in 2014, almost all the respondents (70%) believe that this type of event is useful to promote the role of the researcher and it could encourage the choice of young people to pursue a scientific career (40%). However, 53% of the respondents believe that research and science are not very much adequately funded in Italy. While, they think that they are quite well financed in Europe (64%). The level of participants' satisfaction, regarding the efficiency of the organization, has always been high and constant over the years. Overall, most of the participants has expressed their enthusiasm and congratulated with the organizing team for the professionalism and the choice of speakers. During the European Researchers' Night of 2015, for the second consecutive year, Frascati Scienza collected impressions and suggestions of the participants about the event through the World Café methodology. The objectives of the 2015 edition have been: to collect and show the opinions of the public about research in order to analyse their strengths and weaknesses in promotion and communication, opportunities and criticalities, but above all Frascati Scienza was interested to collect ideas for future editions of the European Researchers' Week/Night in a practical, fun and productive way all together. The roundtable have been held at SAPERmercato in Frascati, the new reinvented space dedicated to science, involving an audience of about 40 people consisting of both adults and children, in a broad debate for identifying interesting ideas. In order to promote a systematic dialogue, a series of keywords have been proposed to participants to help them to focus on the issues.

About opportunities and criticalities Frascati Scienza has presented the following keywords:

- collaborations/partnerships among cities, institutions, associations
- project value as contributions to the "Tuscolana Area";
- the concept of "sustainability"
- multidisciplinary thematic proposals.

The analysis of the opportunities has showed particular appreciation for the event in being able to communicate new themes and engaging people in practical experiments. According to the participants, the European Researchers' Night of 2015 has constituted an opportunity to access to more reliable information since during the event it has been possible to directly speak with researchers. It has also emerged how the event has been useful to promote the Tuscolana area and transmit what has been done from the research world through an inside perspective. About the criticalities, participants have raised doubts on the schedules of events and the fact that sometimes too many activities have been concentrated and they would need more time o participate in. In fact, it could be useful to extend the events on weekends or in other occasions during the year. About the critical issues, some people stated that the programme of the events was unclear and advised to make the website more easy to use.

The World Café also discussed the Strengths and the Weaknesses of the event focusing the discussion on the following keywords:

- general organization
- easiness of registration and access to location/activities
- level of innovativeness of the events
- usability issues
- technical and scientific arguments
- promotion and communication of the event
- involvement and collaboration of the society.

Regarding the Strengths, the participants considered easy to understand the issues proposed. They have enjoyed the open and direct dialogue between the audience and the speakers. Children had the opportunity to talk about many curiosities with researchers. During the debate, some people appreciated the passion of the researchers. The accessibility to the events has been considered adequate and the website excellent. The advertising campaign has been effective and the public felt pampered by Frascati Scienza staff.

About the Weaknesses, some practical aspects emerged, such as the acoustic of some rooms and the need to carry out these activities in larger spaces. On the other hand, some issues related to communication have been raised, such as to anticipate event information in June and in particular the promotion of evens dedicated to schools, giving a proper notice. Not all participants found comfortable starting times of the events and it would be better to postpone them. About organization, it would be desirable to have a pause time after the events and to increase the number of the stands with the experiments considered useful and fun, as well as increasing the number of workshops for younger children (under 6). It also has been asked to organise more events in kindergartens throughout the year. Finally, during the World Café, Frascati Scienza has asked participants to focus on Suggestions for the Future to be taken into account for the organization of the next editions. Also in this case, we have prepared some keywords to focus the attention of participants:

- organization
- reservations and access to the website
- location activities
- type of activities
- interaction with researchers
- future topics to be treated
- any further collaborations/partnerships can be developed
- communication/advertising of the Night event and related activities.

The participants focused on some issues they would have like to go deeper in through laboratories and experiments, such as ecology, chemistry and mathematics. Especially children would like to talk about the Big Bang and they would also increase the number of experiments dedicated to them, which in general are very helpful and fun.

## 5  NEXT STEPS

Frascati Scienza has won the MSCA-NIGHT call funded by the European Commission for both 2016 and 2017 years. Through the "MADE IN SCIENCE" project, Frascati Scienza will organise the European Researchers' Night in Italy and continue its work for promoting education and science communication activities within the society at large. To this end, Frascati Scienza has proposed to focus on the "MADE IN SCIENCE" theme for both the NIGHT 2016 and 2017. As a trademark, we aim to communicate to societies the importance of science in terms of its advantages, such as: quality, identity, creativity, security guarantee, trans-nationality, "Know-how", responsibility. This approach certainly contributes to attract young people to enter a scientific or research career. "MADE IN SCIENCE" will constitute a big challenge for Researchers acting with a primary role for designing and organizing the activities of the NIGHT 2016 and 2017 alongside young students, alternating training to effective job contribution.

"MADE IN SCIENCE" is coordinated by Frascati Scienza and is hosted in the research areas in Italy where more than 5.000 international researchers live: public and laboratories areas of the major Italian cities (Roma, Frascati, Firenze, Sesto Fiorentino, Milano, Trieste, Genova, Modena, Ferrara, Napoli, Caserta, Palermo, Bari, Cagliari, Monserrato, Catania, Lecce, Parma, Pavia, Reggio Emilia, Sassari, Carbonia, Cassino, Gorga, Grottaferrata, Monte Porzio Catone, Colleferro, Rocca di Papa e Santa Maria di Galeria) in collaboration with the following institutions Regione Lazio, Municipality of Frascati, ASI, CNR, ENEA, ESA-ESRIN, INAF, INFN, INGV, ISS, CINECA, GARR, ISPRA, CREA, Sardegna Ricerche, Sapienza Università di Roma, Università degli Studi di Roma "Tor Vergata", Università degli Studi di Roma Tre, Università LUMSA di Roma e Palermo, Università di Cagliari, Università degli Studi di Cassino e del Lazio Meridionale, Università di Parma, Università Politecnica delle Marche, Università di Sassari, Università degli Studi di Modena e Reggio Emilia.

## 6  CONCLUSIONS

The analysis presented in this paper has shown the relevance and the huge impact of the European Researchers' Night organised by Frascati Scienza researchers and citizens since 2006. As a main result, the event has been judged positively on average by about 85% of the participants in all the editions, and about 98% of respondents have stated that would like to participate again in an event of this type. Throughout the different editions of the European Researchers' Night, on average, 96% of respondents have stated that the role of research is *very important for the development of our country and for Europe*. More in detail, in the last edition of 2015, the overall rating of the "European Researchers' Night" is positive for all respondents: the majority of them are surprised by the pleasantly content of their visits (laboratories, equipment, design research), but also on the ability to popularize researchers and their human side. We provide below in Table 1 the lessons learnt in the implementation of the "European researchers' Night" from 2006 to 2015 that can be used for organising education and promotional activities within the science communication context.

Table 1 Lessons learnt from ERN 2006 - 2015 and adopted solutions.

| LESSONS LEARNT | ADOPTED SOLUTIONS |
|---|---|
| the development of peculiar communication languages is essential to reach different categories of public and to attract a number of participants as large as it is possible | the use of all the available media and communication languages to develop a spectacular Event, planned in partnership with professionals |
| the demand of the public to participate to the guided visits at the research centers is always increasing | extensive opening up at national level of the partners venues and consequent enlargement of the activities to be made integrating the visits with several exhibits and edutainment, emotional and spectacular events |
| kids and parents prefer, if it is possible, to participate separately to dedicated activities | set up of dedicated areas for kids, youth and adults |
| people understand better about the role of researchers when they have the opportunity to see what they do in their labs and what are the possible applications of their job for a "better life" | increase the direct contact between researchers and common people |
| the "communication skills" of researchers, is strategic for the full understanding of their role in society | improve the communication skills of researchers through specific meeting planned before the events in partnership with communication experts |
| European and international dimension of the research are of great impact on the public at large | - Involvement of young researchers granted by MSCA<br>- increase the geographic area of influence of the event and the connections between locations and European partners |
| interactive activities are the most appreciated by the public at large | improve the attractiveness and interactivity of the events |
| promotional events and activities during the previous weeks/months of the NIGHT increase the number of attendees at the event | timely increase the number of promotional activities, locations and venues allow to reach the schools before the summer break |
| Schools participate with a relevant number of students and interest, which increase the participation of attendees to the activities of the NIGHT | specific contests and activities devoted to schools have been organized throughout the week of the NIGHT in order to allow students to participate during mornings and evenings in the activities |

**ACKNOWLEDGEMENTS**

We would like to thank all the main research institutions and universities that collaborated with us during these ten years - ASI, CNR, ENEA, ESA-ESRIN, INAF, INFN, INGV, La Sapienza Università di Roma, Università degli Studi di Roma "Tor Vergata" e Università degli Studi di Roma Tre – the main territorial institutions – Regione Lazio and Comune di Frascati, all the associations and other partners who participated in the various editions. We also want to thank all the main collaborators: Giuseppe Mazzitelli, Fabrizio Murtas, Claudia Ceccarelli, Francesca Neroni, Susanna Lo Iacono, Fabio Agostini, Ida Capra, Fabio Capra, Raffaele Giovanditti, Carlo Mancini, Chiara Medini, Giusi Sanzone, Filippo Faccini, Maurizio Roscani and Simona Tiseo. A special thank to Colette Renier, Project Officer from the European Commission, supporting us during these years in promoting our ideas for bringing the research world closer to societies.

Work partially supported by the European Commission for the following projects: COME-IN-2016-FP6-GA-044837, AGORA-2007-FP7-GA-200202, EOS-2008-FP7-GA-228619, SAY-2009-FP7-GA-244954, BEST-2010-FP7-GA-265743, BRAIN-2011-FP7-GA-287442, RESPECT-2012-FP7-GA-316436,